\documentclass[12pt]{article}

\usepackage{fullpage}
\usepackage{amsmath}
\usepackage{amssymb}
\usepackage{epsfig}
\usepackage{fullpage}
\usepackage{subfigure}
\usepackage{feynmf}
\begin{document}


\begin{center}
{\Large \bf  A Possible Arena for Searching New Physics -\\
the $\Gamma (D^0 \to \rho^0 \gamma)/
\Gamma (D^0 \to \omega \gamma) $ Ratio }\\
\vspace{1cm}
{\large \bf S. Fajfer$^{a,b}$,  S. Prelov\v sek$^{a,c}$,
P. Singer$^{d}$ and 
D. Wyler$^{e }$\\}
\vspace{.5cm}

{\it a) J. Stefan Institute, Jamova 39, P. O. Box 3000, 1001 Ljubljana,
Slovenia}
 \vspace{.3cm}

{\it b)
Department of Physics, University of Ljubljana, Jadranska 19, 1000 Ljubljana,
Slovenia}
\vspace{.3cm}

{\it c)
Department of Theoretical Physics, University of Trieste, Strada Costiera 11,  34014 Trieste, Italy}
\vspace{.3cm}

{\it d) Department of Physics, Technion - Israel Institute  of Technology,
Haifa 32000, Israel}
\vspace{.3cm}

{\it e) Institut f\" ur Theoretische Physik, Universit\" at Z\" urich,
Switzerland}

\end{center}

\vspace{0.25cm}

\centerline{\large \bf ABSTRACT}

\vspace{0.2cm}
We propose to investigate flavour changing neutral currents
in the $c \to u \gamma$ transition through the measurement of the
difference between $\Gamma (D^0 \to \rho^0 \gamma)$ and
$\Gamma (D^0 \to \omega \gamma)$.
This is based on the observation that
$D^0 \to (d \bar d) \gamma$ is due to long distance physics while
$D^0 \to ( u \bar u) \gamma$ arises  from the $c \to u \gamma$
transition. The effect of $\rho - \omega$ mixing is included. A difference 
in the decay widths of more than about $30 \%$ would be indicative of 
new physics.


\vspace{2cm}

Despite the remarkable success of the standard model (SM), it is
generally believed that this is an effective theory at present
energies. The lack of explanation for many of the salient
features of SM suggests that one must look for its extension.
At present, there is no clear picture of the form of the SM
extension and the search for physics beyond the SM proceeds
along many alternative paths.

As the couplings of $Z^0$, photon and Higgs boson are flavour diagonal,
flavour changing neutral current (FCNC) transitions are rare in SM since
they can arise at loop level only.
Moreover, FCNC transitions in the up-quark sector (like
$c\to u \gamma$ ) are much rarer than those in the down-quark
sector (like $b \to s \gamma$) as a result of the CKM matrix
elements and masses involved \cite{BGHP}. Accordingly, they could provide an
appropriate ground for the search of new physics effects
\cite{BGM}. In fact, rare charm decays have been frequently marked as
possible sources for the discovery of
new physics \cite{BGM,P}, in view of the smallness
of the short distance SM contributions of these processes.

Hadronic decays like $D \to V \gamma$ [1, 4-9] 
and $D \to V l^+ l^-$ [9-12] have been considered
with the aim of investigating the size of the short distance
(SD) $c \to u \gamma$, $c \to u l^+ l^-$ contribution
to the decay amplitude. However, it turns out that these decays are
largely dominated by long distance (LD) contributions, rendering them
inappropriate for detecting deviations from the SM values of the
basic $c \to u \gamma (l^+ l^-)$ transitions.
An exception is $B_c \to B_u^* \gamma$ since in this
decay both the SD and LD contributions to the branching ratio are
in the $10^{-8}$ range \cite{FPS3}.

In the present letter we suggest a new possibility for the search of the
$c \to u \gamma$ transition, from the determination of the
difference in size of the partial decay widths
$D^0 \to \rho^0 \gamma$ and $D^0 \to \omega \gamma$.
Our method is particularly suitable for  detecting an
enhancement of the $c \to u \gamma$ amplitude coming from new
physics if it increases this amplitude by at least a factor 3-4.
We remark that enhancements of up to a factor 100
have been noted in certain non-minimal\footnote{Non-minimality here means that the universality of the soft breaking terms is not imposed or that additional super-fields are added to MSSM.} 
 realizations of supersymmetry \cite{BGM}.

Next we note that in the radiative decays of $D^0$ mesons, the
$(d\bar d) \gamma$ final state arises primarily as a
result of nonleptonic $W$- exchange $c\bar u\to d\bar d$, being therefore an 
outcome of
LD physics (Fig 1a), while the $(u\bar u) \gamma$ final state
is mainly due to the electro-magnetic
penguin $c \to u \gamma$ transition (Fig. 1b). The $(u\bar u) \gamma$ final 
state receives also a small  contribution form the long distance penguin mechanism 
illustrated in Fig. 1c. The decays
$D^0 \to (\rho, \omega) \gamma$ have been treated in great detail
in Refs. \cite{FS,FPS1,thesis} and it is shown that the LD
contributions completely overshadow the SD contribution in SM.
Although, there is a strong enhancement at the two - loop level
\cite{GHMW}  of the $c \to u \gamma$ amplitude \cite{KP} as a
result of gluonic corrections, the QCD - corrected SM amplitude
\cite{GHMW} is still only a few percent of the LD one \cite{FPS1}.

As usual, we define the isospin eigenstates
\begin{equation}
\label{1}
\omega^{(I = 0)}  =  \frac{1}{{\sqrt 2}} ( \bar u u + \bar d d)~,\qquad
\rho^{(I = 1)}  =  \frac{1}{{\sqrt 2}} ( \bar u u - \bar d d)~.
\end{equation}
Therefore, in the absence of the $\omega -  \rho$ mixing and the LD penguin contribution, the LD rates for $D^0 \to \rho \gamma$ and
 $D^0\to\omega \gamma$  are equal in the SM; a difference must be due
to SD effects. This  is the basis for 
our approach.

The main decay modes of the physical $\rho$, $\omega$ states are
$2 \pi$ and $3 \pi$ respectively. It is well known that the
$\omega^{(I = 0)}$ and $\rho^{(I = 1)} $ states can mix and $\omega$ is
known to decay also to $2 \pi$ states, with a branching ratio
of $Br( \omega \to \pi^+ \pi^-)$ $ = (2.21 \pm 0.30)\%$
\cite{CASO}. 
The physical states denoted
 as $\rho$ and
$\omega$ are related to the states defined in Eq. (\ref{1}) by \cite{mix}
\begin{equation}
 \label{mix}
 \rho  =  \rho^{(I = 1)} - \epsilon ~\omega^{(I = 0)}~,\qquad
 \omega  =  \omega^{(I = 0)} + \epsilon~ \rho^{(I = 1)}~,
\end{equation}
where 
\begin{equation}
\label{8}
 \epsilon  =   \frac{\Pi_{\rho \omega}^2}{
\hat m_{\omega}^2 - \hat m_{\rho}^2}\simeq\frac{\Pi_{\rho \omega}^2}{
m_{\omega}^2 - m_{\rho}^2 + i m_{\rho} \Gamma_{\rho}-im_\omega\Gamma_\omega } 
\end{equation}
with complex masses $\hat m=m+im\Gamma/2$ and with real $m$ and $\Gamma$. 
This  leads to the $\omega \to 2 \pi$ amplitude
\begin{equation}
\label{4}
{\cal A} (\omega \to 2 \pi)  =  \frac{\Pi_{\rho \omega}^2}{
m_{\omega}^2 - m_{\rho}^2 + i m_{\rho} \Gamma_{\rho}-im_\omega\Gamma_\omega } ~
{\cal A} (\rho \to 2 \pi),
\end{equation}
from which the mixing parameter is determined to be
\begin{eqnarray}
\label{5}
 \Pi_{\rho \omega}^2 & = & - (4.0 \pm 0.4) \times 10^{-3}
 \enspace {\rm GeV}^2
\end{eqnarray}
where the values of $m_\rho$, $m_\omega$ and $\Gamma_\rho$ are taken from 
\cite{CASO}. 
The minus sign in (\ref{5}) is obtained from a detailed analysis of
the $e^+ e^- \to \pi^+ \pi^-$ amplitude \cite{mix}. The corresponding value of $\epsilon$ (\ref{8}) is
\begin{eqnarray}
 \label{epsilon}
\enspace  \epsilon & = &-0.0061 + i ~0.036 ,
\end{eqnarray}
and the physical states can be written as 
\begin{equation}
 \label{11}
\rho  = \frac{1}{\sqrt{2}} (1 - \epsilon) u \bar u +
 \frac{1}{\sqrt{2}} (- 1  - \epsilon)d \bar d\qquad 
 \omega  = \frac{1}{\sqrt{2}}  (1 +\epsilon) u \bar u +
 \frac{1}{\sqrt{2}} (1  - \epsilon)d \bar d~
 . 
\end{equation}

The general form of a $D \to V \gamma$ transition amplitude
is written as usual in terms of parity conserving
$A_{PC}$ and parity violating $A_{PV}$ amplitudes  
\cite{BGHP,FS,FPS1}
\begin{align}
 \label{13}
{\cal A}[D^0 (p) \to V(p_V,\epsilon_V) \gamma(q,\epsilon_\gamma)]
 = -i\{& A_{PC} \epsilon_{\mu \nu \alpha \beta} q^\mu
\epsilon_{\gamma}^{*\nu} p^\alpha \epsilon_V^{*\beta}\\
 +& i A_{PV} [ (\epsilon_V^* \cdot q)(\epsilon_\gamma^* \cdot p_V)
  - (q \cdot p_V) (\epsilon_V^* \cdot \epsilon_\gamma^*)]\}.\nonumber
\end{align}
We decompose the $A_{PC}$ and $A_{PV}$ amplitudes according to their
final state content
\begin{equation}
A_{PC}  =   A_{PC}( u \bar u) + A_{PC}(d \bar d),\qquad
A_{PV}  =  A_{PV}( u \bar u) + A_{PV}(d \bar d)~.
\end{equation}
At this point we define the new quantities $\eta_{PC}$, $\eta_{PV}$
\begin{equation}
\label{eta}
\eta_{PC}  =  \frac{A_{PC}( u \bar u)}{ A_{PC}(d \bar d)}~,
\qquad \eta_{PV}  =  \frac{A_{PV}( u \bar u)}{ A_{PV}(d \bar d)}~,
\end{equation}
which, as we explain later on, are of the order of a few percent in the standard model, and the amplitudes are rewritten as
\begin{align}
\label{18}
A_{PC/PV}(D^0 \to\rho^0 \gamma)  &= \frac{1}{{\sqrt 2}} A_{PC/PV}(d\bar d)~
[(1 - \epsilon ) \eta_{PC/PV} + (-1 - \epsilon)]~,\nonumber\\
A_{PC/PV}(D^0 \to \omega \gamma)  &= \frac{1}{{\sqrt 2}} A_{PC/PV}(d\bar d)~
[(1 +\epsilon) \eta_{PC/PV} + (1 - \epsilon)].
\end{align}
The decay width is given by
\begin{equation}
\label{19}
\Gamma(D \to V\gamma)  = \frac{1}{4 \pi}
( \frac{m_D^2 - m_V^2}{2 m_D})^3 (|A_{PC}|^2 + |A_{PV}|^2)~.
\end{equation}
 In order to extract the $(u\bar u)\gamma$ final state, we propose a quantity ${\cal D}^{\omega-\rho}$, defined as
\begin{equation}
\label{dmix}
{\cal D }^{\omega-\rho} \equiv  \frac{ \Gamma [ D^0 \to \omega \gamma] /
(m_D^2 - m_\omega^2)^3 - \Gamma [ D^0 \to \rho^0 \gamma]
/(m_D^2 - m_\rho^2)^3}{\Gamma [ D^0 \to \omega \gamma]
/(m_D^2 - m_\omega^2)^3}~.
\end{equation}
The standard model values of $\eta_{PC/PV}$ (given in (\ref{etasm}) bellow) and
 $\epsilon$ (\ref{epsilon}) are small and ${\cal D }^{\omega-\rho}$ can be 
expanded to the first order
\begin{equation}
\label{21}
{\cal D }^{\omega-\rho} \simeq 4~ \frac{|A_{PC}(d \bar d)|^2~ ({\rm Re}~ 
\eta_{PC}-{\rm Re}~\epsilon) +
|A_{PV}(d \bar d)|^2~ ( {\rm Re}~ \eta_{PV}-{\rm Re}~\epsilon)}
{ |A_{PC}(d \bar d)|^2  +
|A_{PV}(d \bar d)|^2 }~,
\end{equation}
which leads to  the near equality of the $D^0 \to \rho \gamma$ and
 $D^0 \to \omega \gamma$ rates. On the other hand, physics beyond
 the standard model, which affects significantly the size of
 $c \to u \gamma$ transition, will lead to a splitting of the
 $\Gamma [D^0 \to \rho^0 \gamma] \simeq \Gamma [D^0 \to \omega \gamma]$ near 
equality. Sizeable coefficients 
${\cal D}^{\omega-\rho}$  can arise in such scenarios and the expansion 
(\ref{21}) is not valid. We remark that as a
consequence of the mixing term $\epsilon\not = 0$ (\ref{mix})
 there is a 
 difference of a few percent between the widths of the two decays, even when
$\eta_{PC}$ and $\eta_{PV}$ are neglected. An additional difference arises at the hadronic level as a consequence of $SU(3)$ flavour breaking. For example, as one sees from the explicit expressions of \cite{FS}, the difference in couplings and masses of $\rho$ and $\omega$ induces a $\sim 10\%$ difference between the $D^0\to \rho^0\gamma$ and $D^0\to \omega\gamma$ widths.

We estimate the long distance contribution $A(d\bar d)$ from the 
 calculation of the decays $D^0 \to \rho^0\gamma,$ $ \omega \gamma$
derived by the use of the HQET and chiral Lagrangian together with 
factorization in Refs. \cite{FS,FPS1}. The relative signs of different 
contributions to the parity conserving and parity violating amplitudes were 
left undetermined in Refs.   \cite{FS,FPS1} and are taken from the quark models, as in \cite{thesis}, giving
\begin{equation}
\label{dd}
  A_{PC} (d \bar d) \simeq  -3.7 \times 10^{-9}\enspace  {\rm GeV}^{-1}~,
\qquad
  A_{PV} (d \bar d) \simeq  9.5 \times 10^{-9}\enspace  {\rm GeV}^{-1}~.
\end{equation}
 Inserting the numerical values of $A(d\bar d)$ (\ref{dd}) and $\epsilon$ 
(\ref{epsilon}) to the expression for  ${\cal D}^{\omega-\rho}$ (\ref{dmix}) we
 get 
\begin{equation}
\label{dmix1}
{\cal 
D}^{\omega-\rho}\!\simeq\!\frac{0.025+0.54~{\rm Re}\eta_{PC}+3.6~{\rm Re}\eta_{PV}-0.003~|
\eta_{PC}|^2-0.02~ |\eta_{PV}|^2}{1+0.27~{\rm Re}\eta_{PC}+1.8~{\rm Re}\eta_{PV}-0.02~{\rm 
Im}
\eta_{PC}-0.13~{\rm Im}\eta_{PV}+0.14~|\eta_{PC}|^2+0.9~|\eta_{PV}|^2}
\end{equation}
where all orders in  $\eta_{PC/PV}$ are retained. 
This expression displays the sensitivity of the quantity ${\cal 
D}^{\omega-\rho}$, defined by (\ref{dmix}), on the magnitude of the short 
distance rate $c\to u\gamma$ contained in $\eta_{PC/PV}$ (\ref{eta}). We note 
that even if we neglect $\eta_{PC/PV}$, there is a difference of about $2.5\%$ 
in the decay widths due to the $\rho-\omega$ 
mixing (\ref{mix}, \ref{epsilon}), and another possible difference coming from $SU(3)$ breaking; together these amount to at most $15\%$.   

The standard model prediction for the short distance contribution $A^{SD}(u\bar u)$ is extracted from the calculation of the $c\to u\gamma$ rate at the two loop level \cite{GHMW} giving  $Br(c\to u\gamma)\sim 3\times 10^{-8}$ for $D^0$ decays.  The matrix  element
$\langle \rho^0,\omega | \bar u \sigma_{\mu \nu} (1 + \gamma_5) c | D^0\rangle 
$ is calculated using the procedure described in
detail in Ref. \cite{FS}, giving
\begin{eqnarray}
\label{22}
A_{PC}^{SD} (u\bar u) & \simeq & (-1.4 \times 10^{-10} -
i ~4.0 \times 10^{-10})~[1\pm 0.2]\enspace  {\rm GeV}^{-1}~,\nonumber\\
A_{PV}^{SD} (u\bar u) & \simeq & (-2.3 \times 10^{-10} -
i ~6.6 \times 10^{-10})~[1\pm 0.2]\enspace  {\rm GeV}^{-1}~.
\end{eqnarray}
The mechanism, shown in Fig . 1c, gives rise to a long distance part of $A(u\bar  u)$ and has been estimated in \cite{FS,thesis}
\begin{equation}
\label{22a}
A_{PC}^{LD} (u\bar u)  \simeq (2.2 \pm 2.2)\times 10^{-10}\enspace  {\rm 
GeV}^{-1}~,\qquad A_{PV}^{LD} (u\bar u)  \simeq  (3.7 \pm 3.7)\times 
10^{-10}\enspace  {\rm GeV}^{-1}~.
\end{equation} 
This contribution is small as it is proportional to the $SU(3)$ flavour breaking parameter
\begin{equation}
\label{cvmd} 
 {g_{\rho}^2(0)\over  2m_{\rho}^2}-{g_{\omega}^2(0)\over  
6m_{\omega}^2}-{g_{\phi}^2(0)\over  3m_{\phi}^2}\simeq (-1.2\pm1.2)\times 
10^{-3}~{\rm GeV}^2
\end{equation}
with  $g_V$ defined as $\langle 
V(q,\epsilon)|j_V^{\mu}|0\rangle=g_V(q^2)\epsilon^{*\mu}$. The mean value and 
error in (\ref{cvmd}) are determined from the experimental data on $V^0\to 
e^+e^-$ decays \cite{CASO} by assuming $g_{V}(0)\simeq g_{V}(m_{V}^2)$.  Given that $A(u\bar u)=A^{SD}(u\bar u)+A^{LD}(u\bar u)$ (\ref{22}, \ref{22a}), the 
standard model prediction for $\eta_{PC/PV}$ is
\begin{alignat}{2}
\label{etasm}
{\rm Re} ~\eta_{PC}&\simeq -0.02\pm 0.06~&,\qquad {\rm Im}~ \eta_{PC}&\simeq 0.11\pm 
0.02\\
{\rm Re} ~\eta_{PV}&\simeq 0.01\pm 0.04~&,\qquad {\rm Im} ~\eta_{PV}&\simeq -0.07\pm 
0.01\nonumber
\end{alignat}
leading to the approximate equality
\begin{equation}
Br[D^0\to \rho\gamma]\simeq Br[D^0\to \omega\gamma]\simeq 1.2\times 10^{-6}~.
\end{equation}
The standard model prediction for the relative difference of the rates ${\cal D}^{\omega-\rho}$ 
(\ref{dmix}, \ref{dmix1})  is 
\begin{equation}
{\cal D}^{\omega-\rho}_{SM}\simeq 6\pm 15~~\%~, 
\end{equation} 
where the error is dominated by the uncertainty of the $SU(3)$ flavour breaking
 parameter, including  (\ref{cvmd}). The strong rescattering, which can transform a $d \bar d$ pair to a 
$u \bar u$ pair does not change this result. This is because 
we consider decays into strong interaction eigenstates $\rho$ and $\omega$, which evolve as $\exp[-i(E-i\tfrac{1}{2}\Gamma)t]$.

For completeness, we consider also the case, when the $\rho$ and $\omega$  states are experimentally identified only by the $2\pi$ and $3\pi$ decay modes, respectively, and not  by their masses and widths. If the two pion final states, which arise from the decay of $\omega$, are not disentangled in the experiment, then one is dealing with the final states of good isospin. A final state, which starts out as $\rho^{(I=1)}$, will obtain an admixture of  $\omega^{(I=0)}$ due to the isospin mixing (\ref{mix}), and vice versa. The production of $2\pi\gamma$ and $3\pi\gamma$  final states depends on time. In order to extract the short distance contribution, we propose to look at the number of $2\pi\gamma$ and $3\pi\gamma$, integrated over time
 \begin{equation}
\label{d}
{\cal D }^{3\pi-2\pi} \equiv  \frac{\frac{ N [ D^0 \to 3\pi ~\gamma]} {
(m_D^2 - m_\omega^2)^3 Br(\omega\to 3\pi)} - \frac{N [ D^0 \to 2\pi~ \gamma]}{
(m_D^2 - m_\rho^2)^3}}{\frac{N[ D^0 \to 3\pi~ \gamma]}{
(m_D^2 - m_\omega^2)^3Br(\omega\to 3\pi)}}~,
\end{equation}
where $2\pi$ and $3\pi$ have invariant masses covered by $\rho$ and $\omega$ resonances.
At the first order in $\eta_{PC,PV}$ and $\epsilon$  we get in the standard model
\begin{equation}
\label{time}
{\cal D }^{3\pi-2\pi}_{SM} \simeq 4~ \frac{|A_{PC}(d \bar d)|^2~ {\rm Re}~\eta_{PC}
 + |A_{PV}(d \bar d)|^2~  {\rm Re}~ \eta_{PV}}
{ |A_{PC}(d \bar d)|^2  +|A_{PV}(d \bar d)|^2 }\simeq 4\pm 15~~\%~,
\end{equation}
where we have taken into account that $\epsilon(E_{\omega}-E_{\rho})/\Gamma_\rho\ll \eta_{PC,PV}~$.

The quantities ${\cal D}^{\omega-\rho}$ (\ref{dmix}) and ${\cal D }^{3\pi-2\pi} $ (\ref{d}) are particularly sensitive to new physics 
scenarios, which could enhance the $c\to u\gamma$ rate. Non-minimal 
realizations of the supersymmetric  models, discussed in \cite{BGM}, can enhance 
these quantities up to  
\begin{equation}
  {\cal D}^{\omega-\rho}\simeq  {\cal D}^{\omega^0-\rho^1}\simeq 1 ~.
\end{equation}  

\vspace{0.2cm}

Similar reasoning may be useful in extracting the FCNC transition $c\to ul^+l^-
$ from the difference in the decay rates $\Gamma[D^0\!\to\! \rho^0l^+l^-]$ and 
$\Gamma[D^0\!\to\! \omega l^+l^-]$, where $l$ denotes an electron or muon. Long distance contributions to these  have been studied in \cite{thesis,FPS2} 
and arise via the mechanisms illustrated in Figs. 1a and 1c, where the photon 
is replaced with the virtual photon decaying to leptons. The corresponding 
differential branching ratio has resonant shape in terms of the di-lepton mass 
$m_{ll}$ with maximums at $m_{ll}=m_\rho$, $m_\omega$ and $m_\phi$. The LD 
contributions, which give rise to the final state $(d\bar d)l^+l^-$,  largely 
cancel in the difference  $\Gamma[D^0\to \omega^0l^+l^-]-\Gamma[D^0\to \rho 
l^+l^-]$, which is proportional to $A(u\bar u)=A^{SD}(u\bar u)+A^{LD}(u\bar u)$
. In this case, however, the  $SU(3)$ flavour cancellation in $A^{LD}(u\bar u)$ is not so effective as the maximums at $m_{ll}\!=\!m_{\rho,\omega}$ and 
$m_{ll}\!=\!m_{\phi}$ are well separated\footnote{The standard model predictions presented in \cite{thesis,FPS2} give  $Br^{LD}(u\bar u)\sim  10^{-7}$ and $Br^{SD}(u\bar u)\sim  10^{-9}$.}. As a consequence, $|A^{LD}(u\bar u)|$ is 
more than one order of magnitude larger than  $|A^{SD}(u\bar u)|$   and  overshadows the 
interesting FCNC transition $c\to ul^+l^-$ in  the difference  $\Gamma[D^0\to 
\omega^0l^+l^-]-\Gamma[D^0\to \rho l^+l^-]$.

\vspace{0.2cm}

To summarize, we have proposed here yet another test for physics beyond the 
standard model in the charm sector. Our test exploits the extreme smallness 
of the $c\to u\gamma$ transition in standard model \cite{GHMW} and the near 
equality of the $D\to \rho^0\gamma$, $D^0\to\omega\gamma$ amplitudes, 
obtained from the calculation \cite{FS,FPS1,thesis} of long distance contributions 
to these decays. This equality would be spoiled, if, as encountered in certain 
supersymmetric models \cite{BGM}, the $c\to u\gamma$ amplitude is enhanced by 
a seizable factor. Standard model calculations, including $SU(3)$ breaking in 
the long distance contribution to the amplitudes, show that  
$D\to \rho^0\gamma,~\omega\gamma$   widths do not differ by 
more than $15\%$. Thus, we are led to claim, conservatively, that a 
difference larger than $30\%$ would be a ``smoking gun'' indication of 
new physics. The formalism leading to this conclusion has been exposed 
in this paper. In view of the expected $\sim 10^{-6}$ branching ratio 
for these decays, we look forward to the experimental tests, keeping 
in mind that the present upper limits are around $ 10^{-4}$ 
\cite{cleo.radiative}.

\begin{figure}[h]
\begin{center}
\includegraphics[scale=.9]{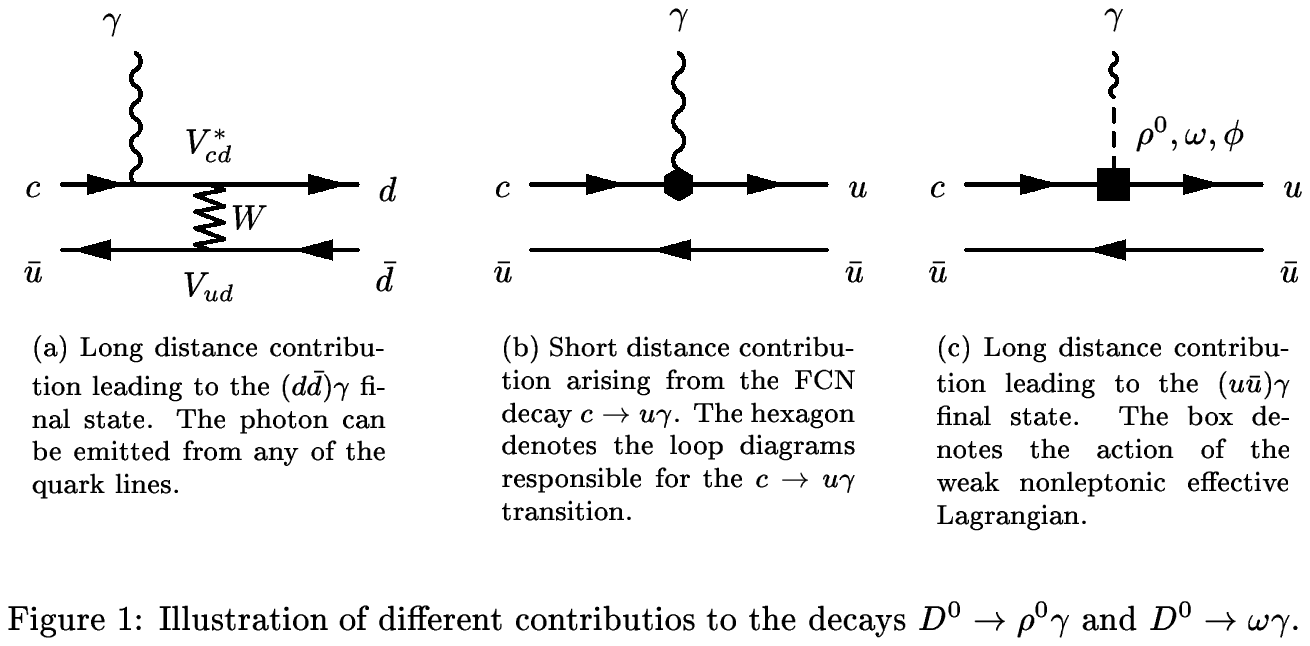} 
 \caption{} 
\end{center}
\end{figure}

\end{document}